\documentclass[fleqn,10pt]{wlscirep}
\usepackage{epsfig}
\usepackage{subfigure}
\usepackage{amssymb,amsmath}
\usepackage{graphicx}
\usepackage{epstopdf}
\usepackage{epsfig}
\usepackage{color}
\usepackage{amsfonts}
\usepackage{amscd}
\usepackage{hyperref}
\usepackage{bbm}

\title{Spin-orbit coupling and electric-dipole spin resonance in a nanowire double quantum dot}

\author[1]{Zhi-Hai Liu }

\author[1]{Rui Li}

\author[2,*]{Xuedong Hu}

\author[1,*]{J. Q. You}

\affil[1]{Quantum Physics and Quantum Information Division, Beijing Computational Science Research Center, Beijing 100193, China}
\affil[2]{Department of Physics, University at Buffalo, SUNY, Buffalo, New York 14260-1500, USA}
\affil[*] {Correspondence and requests for materials should be addressed to J.Q.Y. (email: jqyou@csrc.ac.cn) or X.H. (email: xhu@buffalo.edu)}

\begin{abstract}
We study the electric-dipole transitions for a single electron in a double quantum dot located in a semiconductor nanowire. Enabled by spin-orbit coupling (SOC), electric-dipole spin resonance (EDSR) for such an electron can be generated via two mechanisms: the SOC-induced intradot pseudospin states mixing and the interdot spin-flipped tunneling. The EDSR frequency and strength are determined by these mechanisms together. For both mechanisms the electric-dipole transition rates are strongly dependent on the external magnetic field.  Their competition can be revealed by increasing the magnetic field and/or the interdot distance for the double dot. To clarify whether the strong SOC significantly impact the electron state coherence, we also calculate relaxations from excited levels via phonon emission.  We show that spin-flip relaxations can be effectively suppressed by the phonon bottleneck effect even at relatively low magnetic fields because of the very large $g$-factor of strong SOC materials such as InSb.
\end{abstract}

\begin{document}

\flushbottom
\maketitle

\thispagestyle{empty}

\section*{Introduction}

Confined electron spins in semiconductor nanostructures are a viable option for implementing quantum computing and quantum information processing because of their long decoherence times~\cite{Fabian2004,Hanson2007,Wolf2001,Awschalom1999,Witzel2007,Cywinski2009,Hu2006}, and quantum coherent manipulation of a single electron spin is an essential ingredient for such applications. Conventional approach for manipulating an electron spin uses magnetic dipole interaction to achieve electron spin resonance (ESR)~\cite{Slichter1980}.  However, the very small electron spin magnetic moment dictates that a strong alternating-current (AC) magnetic field is required to reach reasonable rate of spin rotation~\cite{Koppens2006, Press2008}.  In semiconductors, interestingly, spin-orbit coupling (SOC) offers a viable alternative.  Through SOC an AC electric field can also rotate an electron spin, leading to the so-called electric-dipole spin resonance (EDSR)~\cite{ Golovach2006, Nowack2007, Borhani2012, Pfund2007,  Rashba2003,Osika2013, Hu2012,Fabian2008}. Indeed, EDSR has proven to be an effective method for electron spin control in quantum dots~\cite{ Perge2012, Perge2010, Petersson2012}.

Over the past decade semiconductor nanowire devices have attracted wide attention because of their one-dimensionality, convenience of growth, and a variety of interesting physical properties~\cite{Loss2008, Nowak2013, Perge2012, Perge2010, Petersson2012, Deng2016, Li2017, Erlingsson2010, Marques2005, Villegas-Lelovsky2009, Dias2014}.  Experimentally, electron occupancy of quantum dots in a nanowire can be effectively controlled by regulating the local gate electrodes ~\cite{Fuhrer2007,Fasth2005,Fuhrer2007I,Jung2012}. Recently, nanowires with narrow bandgap, large SOC, and large $g$-factor have been of particular interest because they present intriguing opportunities for studying fast electrical control of spins \cite{Petersson2012, Perge2010, Perge2012, Marques2005, Villegas-Lelovsky2009, Dias2014}, possible manipulation of entangled spins \cite{Egues2002, Egues2005}, and hybrid structures made of a superconductor and a large-SOC nanowire are a promising system to search for Majorana fermions \cite{Deng2016, Li2017}.

A double quantum dot (DQD) is an interesting physical system that has attracted considerable attention over the past two decades \cite{Hanson2007}.  The tunnel coupling between two dots significantly alters the energy spectrum of the system as compared to a single dot, which allows fundamentally and technologically important phenomena such as Pauli spin blockade \cite{Hanson2007,Ono2002}.  Another example is the recent demonstration of strong spin-photon coupling in a double dot, where the DQD energy spectrum plays a crucial role in enhancing the spin-photon coupling strength \cite{Hu2012,Mi2017}.

In this paper, we investigate the electronic properties of a nanowire double quantum dot, with a particular focus on the interplay between SOC and the DQD potential on the electric-dipole transitions of a single confined electron.  We obtain the low-energy spectrum of a single electron in the DQD using the linear combination of atomic orbital (LCAO) method~\cite{Stano2005,Hu2000,Yang2011}.  In our calculation the single-dot single-electron orbitals are obtained by accounting for the spin-orbit coupling exactly while treating the external magnetic field as a perturbation \cite{Ruili2013}. In the presence of an alternating electric field applied along the wire axis, EDSR can be generated by spin state hybridization from SOC. In a single or isolated QD, the state hybridization originates from the SOC-induced intradot orbital states mixing.  In a DQD, on the other hand, interdot tunneling can also contribute to orbital mixings. Thus, in a DQD there are two mechanisms leading to the EDSR, and the dominant mechanism can be altered by changing system parameters. When orbital mixing is dominated by the interdot tunneling, we examine how the electric-dipole transition rates depend on the magnitude and orientation of the applied magnetic field. The competition between contributions from the intradot and interdot orbital mixings can be revealed in the variations of the EDSR frequency with the magnetic field strength, at a large interdot distance.  More specifically, we show that at lower applied magnetic field, spin flip assisted by interdot tunneling makes the dominant contribution to EDSR. With increasing the interdot distance and the associated suppression of tunneling, the main mechanism of EDSR in a DQD changes from the interdot spin-flipped tunneling to the intradot orbital states mixing. Finally, we calculate the rates of phonon-assisted spin relaxation and show that the enhancement in relaxation would not significantly impact the quantum coherence quality factor of the electron spin. This study provides useful input for experimental studies of quantum coherent manipulations in a nanowire DQD.

\section*{Results}

\section{The model Hamiltonian}
\begin{figure}[!ht]
\centering
\includegraphics[width=0.55\textwidth]{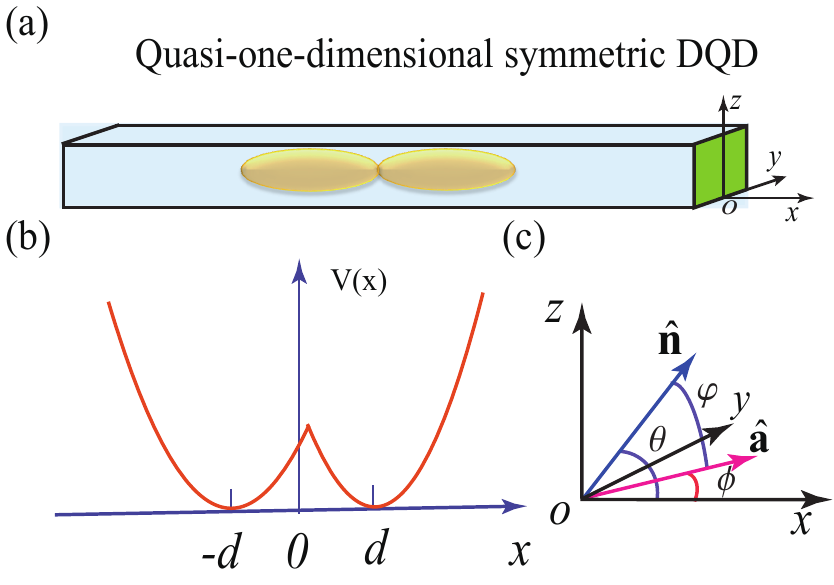}
\caption{(color online) (a) Schematic diagram of a quasi-one-dimensional symmetric DQD in a nanowire. (b) The double-harmonic confinement potential along the interdot axis ($x$ axis), with $2d$ being the interdot distance. (c) Schematic diagram of the unit vectors $\hat{\mathbf{a}}$ and $\hat{\mathbf{n}}$, where $\hat{\mathbf{n}}=(\cos\theta,0,\sin\theta)$ gives the external magnetic field direction, and $\hat{\mathbf{a}}=(\cos\phi,\sin\phi,0)$ is the SOC-induced effective field direction, with $\phi=\arctan(\alpha_{R}/\alpha_{D})\in [0, \pi/2]$ characterizing the relative strength between the Rashba and Dresselhaus SOCs. $\varphi$ is the angle between vectors $\hat{\mathbf{a}}$ and $\hat{\mathbf{n}}$, i.e., $\varphi=\arccos\langle\hat{\mathbf{a}}\cdot\hat{\mathbf{n}}\rangle$. }
\label{model}
\end{figure}

We consider a quasi-one-dimensional double quantum dot with one confined electron, as shown in Fig.~\ref{model}(a).  The semiconductor materials for the nanowires we consider are those with large SOC, such as InAs and InSb~\cite{Romano2005, Sousa2003}, though our approach is sufficiently general so that our results should be applicable to material systems with weaker SOC as well.  To better model a realistic nanowire DQD, we consider an asymmetric nanowire DQD, with system parameters taken from the experimental data of Nadj-Perge et al. in Ref.~\citeonline{Perge2012}.

As illustrated in Fig.\ref{model}(a), the nanowire axis is along the $x$-direction.  Along the transverse directions, we have a strong harmonic potential along the $y$-direction and an asymmetric gradient potential along the $z$-direction (used to enhance the Rashba SOC~\cite{Sousa2003}).  With DQD confinement potential much weaker than the $y$ and $z$ confinements, we treat our electron as quasi-one-dimensional.

In the absence of an applied magnetic field, the Hamiltonian describing an electron in a quasi-one-dimensional DQD along the $x$-direction is
\begin{align}
H_{0}=\frac{p_{x}^{2}}{2m_{e}}+V(x)+H^{x}_{\rm so},
\label{ori}
\end{align}
where $m_{e}$ is the conduction-band effective mass, and $p_{x}=-i\hbar\partial/\partial x$. We choose to model the confinement potential along the $x$ direction as an asymmetric double-well potential $V(x)=\frac{1}{2}m_{e}\min\{\omega_{l}^{2}(x+d)^{2},\omega_{r}^{2}(x-d)^{2}\}$, with $2d$ being the interdot distance and $x_{l/r}=\sqrt{\hbar/(m_{e}\omega_{l/r})}$ being a characteristic length in the left/right dot [see Fig.~\ref{model}(b)]. $H^{x}_{\rm so}$ corresponds to the effective SOC Hamiltonian along the axis direction of the nanowire DQD.

There are two kinds of spin-orbit interactions in $\rm A_{III}B_{V}$  heterostructures~\cite{Winkler2003}. One is the Dresselhaus SOC due to bulk inversion asymmetry~\cite{Dresselhaus1955}. The other is the Rashba SOC generated by structure inversion asymmetry~\cite{Rashba1984}. In general, the SOC strengths depend on system parameters and spatial distributions of the electron wave function. By averaging over the transverse directions $y$ and $z$, we obtain an effective linear SOC Hamiltonian $H^{x}_{\rm so}$ along the $x$ direction (see Methods)
\begin{align}
H^{x}_{\rm so}=\alpha p_{x}\sigma^{a},
\end{align}
with the effective SOC strength $\alpha=\sqrt{\alpha^{2}_{\rm D}+\alpha^{2}_{\rm R}}$.  Here the spin quantization axis is defined by the SOC to be along $\hat{\mathbf{a}}=(\cos\phi,\sin\phi,0)$, with $\phi=\arctan(\alpha_{R}/\alpha_{D})$, so that $\sigma^{a}=\hat{\mathbf{a}}\cdot\boldsymbol{\sigma}$,
where $\boldsymbol{\sigma}=(\sigma_{x},\sigma_{y},\sigma_{z})$ are the Pauli matrices.  $\alpha_{\rm R}$ and $\alpha_{\rm D}$ denote the effective strengths of the Rashba and Dresselhaus SOCs, respectively.

When an external magnetic field is applied in the direction $\hat{\mathbf{n}}=(\cos\theta,0,\sin\theta)$ with strength $B$, the single-electron Hamiltonian becomes
\begin{align}
H'_{\rm DQD}=\frac{(p_{x}+eA_{x})^{2}}{2m_{e}}+V(x)+\alpha p_{x}\sigma^{a}+\frac{g_{e}\mu_{B}B}{2}\sigma^{n},
\label{Hami}
\end{align}
with the vector potential $\mathbf{A} = (A_x, A_y, 0)$, where $A_{x} = -By\sin\theta$, $A_{y} =-Bz\cos\theta $, $g_{e}$ is the Land\'{e} factor, $\mu_{B}$ is the Bohr magneton, and $\sigma^{n}=\hat{\mathbf{n}}\cdot\boldsymbol{\sigma}$. With our assumption of an asymmetric double dot, it follows naturally that the specific value of the Land\'{e} factor $g_{e}$ in the left dot is different from that in the right dot, $g_{el}\neq g_{er}$~\cite{Perge2012}. Due to the strong confinements along the transverse directions, $\langle y \rangle \sim 0$, the effects of the magnetic vector potential on the electron orbital dynamics is negligible (detailed calculations are given in Methods), so that the Hamiltonian for the DQD can be simplified as
\begin{align}
H_{\rm DQD}=\frac{p_{x}^{2}}{2m_{e}}+V(x)+\alpha p_{x}\sigma^{a}-\frac{\Delta_{Z}}{2}\sigma^{n},
\label{Ham}
\end{align}
with the Zeeman splitting $\Delta_{Z}=-g_{e}\mu_{B}B$.

Traditionally SOC is treated as a perturbation in theoretical calculations for semiconductors. However, such a perturbative approach becomes problematic when SOC is strong, in materials such as InSb~\cite{Ruili2013}. For a comprehensive study of the effect of a strong SOC on the electric-dipole transition in a nanowire DQD, in the following calculations we take the SOC term into consideration precisely while treating the Zeeman term perturbatively.

\section{ Energy spectrum of the DQD}

The energy spectrum of the DQD is calculated by adopting the linear combination of atomic orbitals (LCAO) method. The localized electron wavefunctions are derived by solving the eigenstates of the individual quantum dots. The orthonormal bases used to project the DQD Hamiltonian are obtained by the Schmidt orthogonalization of the local wavefunctions.

Near each of the minima of the DQD potential well along the nanowire axis, $V(x)$ can be approximated as parabolic, $V_{l/r}(x) = \frac{1}{2} m_{e} \omega_{l/r}^{2} (x\pm d)^{2}$. Including the SOC effect, the local Hamiltonian for each single quantum dot can be written as
\begin{align}
H'_{l/r}=\frac{p_{x}^{2}}{2m_{e}}+\frac{1}{2}m_{e}\omega_{l/r}^{2}(x\pm d)^{2}+\alpha p_{x}\sigma^{a},
\end{align}
which is isomorphic to the single dot Hamiltonian $H_{0}$ in Refs.~\citeonline{Ruili2013} and \citeonline{Levitov2003}.

The eigenstates of $H'_{l/r}$ can be solved analytically. Let $|\Phi_{\kappa n\sigma}\rangle$ denote the eigenstates of $H'_{\kappa}$, with orbital quantum number $n=0,1,2,3,...$, $\kappa=l,r$ corresponding to the different quantum dots, and $\sigma=\uparrow,\downarrow$ denoting the electron spin states. Explicitly, $|\Phi_{l/r n\uparrow}\rangle$ and $|\Phi_{l/r n\uparrow}\rangle$ take the form
\begin{align}
|\Phi_{l/r n \downarrow}\rangle=&e^{i(x\pm d)/x_{\rm so}}\psi_{l/r n}(x\pm d)|\downarrow_{a}\rangle,\nonumber\\
|\Phi_{l/r n\uparrow}\rangle=&e^{-i(x\pm d)/x_{\rm so}}\psi_{l/r n}(x\pm d)|\uparrow_{a}\rangle,
\label{ow}
\end{align}
where $\psi_{\kappa n}(x)$  represents an eigenstate of a harmonic oscillator with eigenvalue $(n+1/2)\hbar\omega_{\kappa}$, $x_{\rm so}$ is the effective SOC length $x_{\rm so}=\hbar/(m_{e}\alpha)$, and $|\uparrow_{a}\rangle$ and $|\downarrow_{a}\rangle$ denote the eigenstates of $\sigma^{a}$: $\sigma^{a}|\uparrow_{a}\rangle=|\uparrow_{a}\rangle$ and $\sigma^{a}|\downarrow_{a}\rangle=-|\downarrow_{a}\rangle$. $|\Phi_{\kappa n\uparrow}\rangle$ and $|\Phi_{\kappa n\downarrow}\rangle$ are degenerate (Kramers degeneracy), with the eigenvalue given by $\varepsilon_{\kappa n} = (n+1/2) \hbar \omega_{\kappa} - (1/2) m_{e}\alpha^{2}$. The energy levels of $H'_{\kappa}$ are thus evenly spaced, with an energy splitting $\Delta_{\kappa S}=\hbar\omega_{\kappa}$.

In the presence of an applied magnetic field, the single-dot Hamiltonian becomes
\begin{align}
H_{\kappa}=H'_{\kappa}-\frac{\Delta_{\kappa Z}}{2}\sigma^{n},
 \end{align}
 where $\Delta_{\kappa Z}$ corresponds to the Zeeman splitting in $\kappa$ dot.
The Zeeman term can be regarded as a perturbation if the ratio
\begin{align}
\xi_{\kappa}\equiv\Delta_{\kappa Z}/\Delta_{\kappa S}\ll 1,
\label{per-b}
\end{align}
i.e. the Zeeman splitting $\Delta_{\kappa Z}$ is much smaller than the orbital splitting $\Delta_{\kappa S}$, dictating a relatively small magnetic field (see the estimate in Ref.~\citeonline{Ruili2013}). Within first-order perturbation theory, the two lowest-energy eigenstates of $H_{\kappa}$ are
\begin{align}
|\Psi^{\pm}_{\kappa}\rangle=&c^{\pm}_{\kappa}|\Phi_{\kappa 0\uparrow}\rangle+d^{\pm}_{\kappa}|\Phi_{ \kappa 0\downarrow}\rangle+i(\xi/2)e^{-\eta^{2}}\sin\varphi \sum^{\infty}_{n=1}\frac{(\sqrt{2}\eta)^{n}}{n\sqrt{n!}}[(-i)^{n}c^{\pm}_{\kappa}|\Phi_{\kappa n \downarrow}\rangle-d^{\pm}_{\kappa}|\Phi_{\kappa n \uparrow}\rangle],
\label{mix}
\end{align}
where
\begin{align}
&c^{\pm}_{\kappa}=\frac{\cos\varphi\pm f_{\kappa}}{\sqrt{2(f_{\kappa}^{2}\pm f_{\kappa}\cos\varphi)}}, \nonumber \\
&d^{\pm}_{\kappa}=\frac{-ie^{-\eta_{\kappa}^{2}}\sin\varphi}{\sqrt{2(f_{\kappa}^{2}\pm f_{\kappa}\cos\varphi)}}, \label{mix2} \\
&f_{\kappa}=\sqrt{\cos^{2}\varphi+e^{-2\eta_{\kappa}^{2}}\sin^{2}\varphi}\,. \nonumber
\end{align}
Here $\varphi=\arccos\langle\hat{\mathbf{a}}\cdot\hat{\mathbf{n}}\rangle$ is the angle between unit vectors $\hat{a}$ and $\hat{n}$ (i.e., the angle between the effective field from SOC and the applied magnetic field), and $\eta_{\kappa} = \sqrt{m_{e} / (\hbar\omega_{\kappa})} \alpha$. It is a ratio between the effective dot size $x_{\kappa}$ and SOC length $x_{\rm so}$, therefore is a measure of the SOC strength relative to the confinement energy. For a nanowire quantum dot, $\eta_{\kappa}$ is generally small, $\eta_{\kappa} \equiv x_{\kappa}/x_{\rm so}\ll 1$, even for materials with strong SOC.  According to Eq.(\ref{mix}), an applied magnetic field generally leads to hybridization of different spin-orbit states in $|\Psi^{\pm}_{\kappa}\rangle$, with the degree of orbital mixing proportional to $\xi_{\kappa} \eta_{\kappa}^{n}e^{-\eta_{\kappa}^{2}}$.

The orbital states localized in different quantum dots are not orthogonal in general. Nevertheless, from the four lowest-energy localized states $|\Psi^{\pm}_{\kappa}\rangle$ ($\kappa=l,r$) and using Schmidt orthogonalization method, we can construct local orthonormal basis states $|\Psi_{l\Uparrow}\rangle$, $|\Psi_{l\Downarrow}\rangle$, $|\Psi_{r\Uparrow}\rangle$, and $|\Psi_{r\Downarrow}\rangle$. Here $\Uparrow$ and $\Downarrow$ refer to the two pseudo-spin states, whose compositions have been modified by the applied magnetic field as compared to the zero-field Kramers degenerate pair.  The analytical expressions for the bases are given in Methods.

Projecting the Hamiltonian $H_{\rm DQD}$ onto this orthonormal basis, the low-energy part of the Hamiltonian $H_{\rm DQD}$ can be written as
\begin{equation}
H_{\rm DQD}=\left(\begin{matrix}
\varepsilon_{l\Uparrow}& 0 & t_{\Uparrow\Uparrow}&t_{\Uparrow\Downarrow}\\
0& \varepsilon_{l\Downarrow} & t_{\Downarrow\Uparrow} & t_{\Downarrow\Downarrow} \\
t^{\ast}_{\Uparrow\Uparrow} &t^{\ast}_{\Downarrow\Uparrow} &\varepsilon_{r\Uparrow}& 0 \\
t^{\ast}_{\Uparrow\Downarrow} &t^{\ast}_{\Downarrow\Downarrow}&0&\varepsilon_{r\Downarrow}
\end{matrix}\right),
\label{dham}
\end{equation}
with
\begin{align}
& \varepsilon_{\kappa\Uparrow}=\langle\Psi_{\kappa\Uparrow}|H_{\rm DQD}|\Psi_{\kappa\Uparrow}\rangle,~~
\varepsilon_{\kappa\Downarrow}=\langle\Psi_{\kappa\Downarrow}|H_{\rm DQD}|\Psi_{\kappa\Downarrow}\rangle,\nonumber\\
&t_{\Uparrow\Uparrow}=\langle\Psi_{l\Uparrow}|H_{\rm DQD}|\Psi_{r\Uparrow}\rangle,~~~~
t_{\Downarrow\Downarrow}=\langle\Psi_{l\Downarrow}|H_{\rm DQD}|\Psi_{r\Downarrow}\rangle, \nonumber \\
&t_{\Uparrow\Downarrow}=\langle\Psi_{l\Uparrow}|H_{\rm DQD}|\Psi_{r\Downarrow}\rangle,~~~~
t_{\Downarrow\Uparrow}=\langle\Psi_{l\Downarrow}|H_{\rm DQD}|\Psi_{r\Uparrow}\rangle \,.
\end{align}
Here $t_{\sigma\sigma}$ ($\sigma=\Uparrow,\Downarrow$) is spin-conserved tunnel coupling, $t_{\sigma\bar{\sigma}}$ is spin-flipped tunnel coupling and $\varepsilon_{\kappa\sigma}$ is the corresponding single-dot energy.  These matrix elements can be obtained by dividing the original Hamiltonian $H_{\rm DQD}$ in Eq.~(\ref{Ham}) into two parts $H_{\rm DQD}=H_{\kappa}+\Delta_{\kappa} V(x)$, with $H_{\kappa}$ either one of the single-dot Hamiltonian, and $\Delta_{\kappa} V(x)=V(x)-V_{\kappa}(x)$ the double dot correction on $H_{\kappa}$. Due to the orthogonality of $|\Psi_{\kappa\sigma}\rangle$, the tunnelings can be calculated as $t_{\sigma\sigma'}=\langle\Psi_{l\sigma}|\Delta_{l} V(x)|\Psi_{r\sigma'}\rangle$, with its magnitude proportional to the interdot wave function overlap, $t_{\sigma\sigma'}\propto \exp(-d^{2}/\bar{x}^{2})$  where $\bar{x}^{2}=(x^{2}_{l}+x^{2}_{r})/2$.

\begin{figure}[!ht]
\centering
\includegraphics[width=0.45\textwidth]{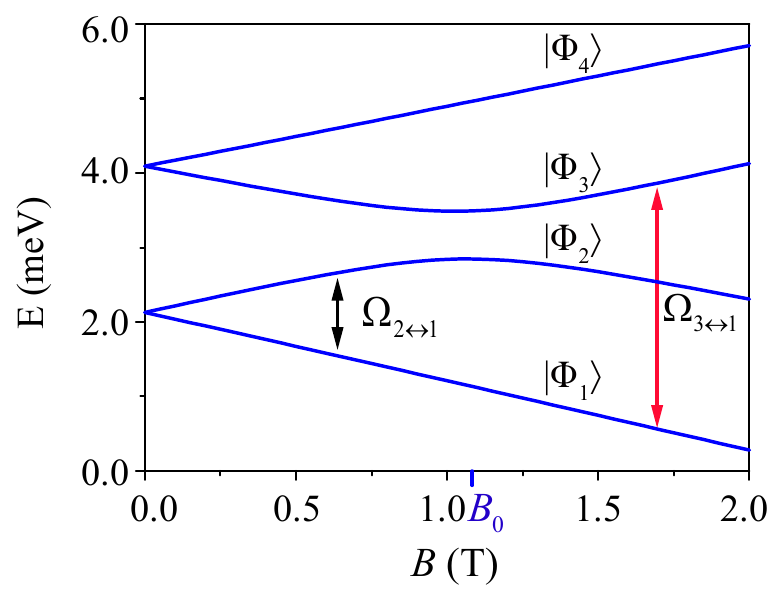}
\caption{(color online) Energy spectrum of the InSb nanowire DQD as a function of the magnetic field strength $B$, with $\varphi=0.5\pi$. The double headed arrows indicate the electric-dipole transitions that we focus on. }
\label{enerb}
\end{figure}

The eigenstates of the nanowire DQD can be obtained numerically by the direct diagonalization of the Hamiltonian $H_{\rm DQD}$ in Eq.(\ref{dham}).  We denote these states $|\Phi_{i}\rangle$ ($i=1-4$), with eigenvalues $E_{1}\leq E_{2}\leq E_{3}\leq E_{4}$. In Fig.~\ref{enerb} we give an example energy spectrum of an InSb nanowire DQD, with the corresponding system parameters taken from the experimental data in Ref.\citeonline{Perge2012}: $\hbar\omega_{l}=5.0~\rm meV$, $\hbar\omega_{r}=7.5~\rm meV$, $g_{el}=-32.2$, $g_{er}=-29.7$,   $x_{\rm so}\simeq 200~\rm nm$, and $d=40~\rm nm$.  The effects of the anisotropic g-factors are neglected for simplicity. Except for the interdot distance and the magnetic field strength and orientation, the parameters of the typical InSb nanowire are used in the following calculations for convenience and consistency.

Equation (\ref{mix}) indicates that when $\xi_{\kappa}\eta_{\kappa} e^{-\eta_{\kappa}^{2}}\ll e^{-d^{2}/\bar{x}^{2}}\ll1$ ($\kappa=l,r$), which is satisfied with the parameters used in Fig.\ref{enerb}, the intradot orbital states hybridization is negligible compared with the interdot states mixing. For weaker SOC or strongly coupled DQD, $x_{\rm so}\gg 2d$, the interdot spin-flipped tunneling $t_{\sigma\bar{\sigma}}$ is much smaller than the spin-conserved tunneling $t_{\sigma\sigma}$. Nevertheless, spin-flipped tunneling leads to a high degree of pseudospin hybridization in states $|\Phi_{2}\rangle$ and $|\Phi_{3}\rangle$ around the anti-crossing point $B_{0}$, as shown in Fig.\ref{enerb}.

In calculating the energy spectrum of the DQD, it is necessary to establish the validity of the perturbation expansion in Eq.(\ref{mix}) and the approximation to neglect orbital effects of the magnetic vector potential in our calculations. The perturbation expansion in Eq.~(\ref{mix}) can be justified by the specific values of $\xi_{\kappa}$ ($\kappa=l,r$) at the upper limit of the magnetic field range we consider. With our chosen parameters, when $B=2$~T, $\xi_{r}<\xi_{l}<0.75$, which still (barely) satisfy the perturbation condition in Eq.(\ref{per-b}).  As for the orbital effect of the vector potential, we compare the effective magnetic length $l_{\rm B} = \sqrt{\hbar / m_{e} \omega_{\rm B}}$, where $\omega_{\rm B} = eB_{\rm t} / m_{e}$ is the electron Larmor frequency, with the characteristic lengths along the transverse directions, with the specific values of $y_{0}$ and $z_{0}$ given in Methods.  At a magnetic field $B=2.0$ T, $l_{\rm B} \simeq 17.78$ nm, which is still larger than the characteristic lengths $y_{0}$ and $z_{0}$. This relationship thus holds true for all the other (lower) fields in our considered parameter regime. Therefore, the approximations we have adopted here are valid in our calculations.

\section{Electric-dipole transitions}

In the absence of SOC, electric-dipole (e-d) interaction induced transitions obey a strict spin selection rule.  In the presence of the SOC, on the other hand, an electric-dipole transition can involve spin flip, leading to EDSR~\cite{Golovach2006, Nowack2007, Borhani2012, Pfund2007, Rashba2003,Osika2013, Hu2012,Fabian2008}. In a DQD with strong SOC, the pseudo-spin composition of the eigenstates vary with magnetic field and interdot distance/tunneling.  Moreover, under certain circumstances, intradot spin mixing in the DQD can also affect EDSR.  In this Section we investigate how EDSR transition rates depend on different system parameters.

When an AC electric field is applied in the $x$ direction, the Hamiltonian describing the single electron in the DQD reads
\begin{align}
H_{\rm e-d}=H_{\rm DQD}+eEx\cos(2\pi\upsilon t),
\end{align}
with $E$ and $\upsilon$ representing the amplitude and frequency of the electric field, respectively. The electric-dipole interaction can be treated as a perturbation if $2eEd \ll \Delta_{ij} \equiv |E_{i}-E_{j}|$, and the resonant electric-dipole transition rate can be calculated as
\begin{align}
\Omega_{i\leftrightarrow j}=(eE/h)\langle\Phi_{i}|x|\Phi_{j}\rangle,
\end{align}
where $h$ is the Plank constant. Due to the spinless e-d interaction, the compositions of the pseudospin states $|\Phi_{i}\rangle$ and $|\Phi_{j} \rangle$ are a crucial factor in determining the magnitude of $\Omega_{i\leftrightarrow j}$.  With the transitions involving state $|\Phi_{4} \rangle$ symmetric with respect to those involving state $|\Phi_{1} \rangle$, for simplicity we only consider the electric-dipole transitions involving $|\Phi_{1} \rangle$ in the following calculations.

\subsection{Magnetic field dependence}

\begin{figure}[!ht]
\centering
\includegraphics[width=0.7\textwidth]{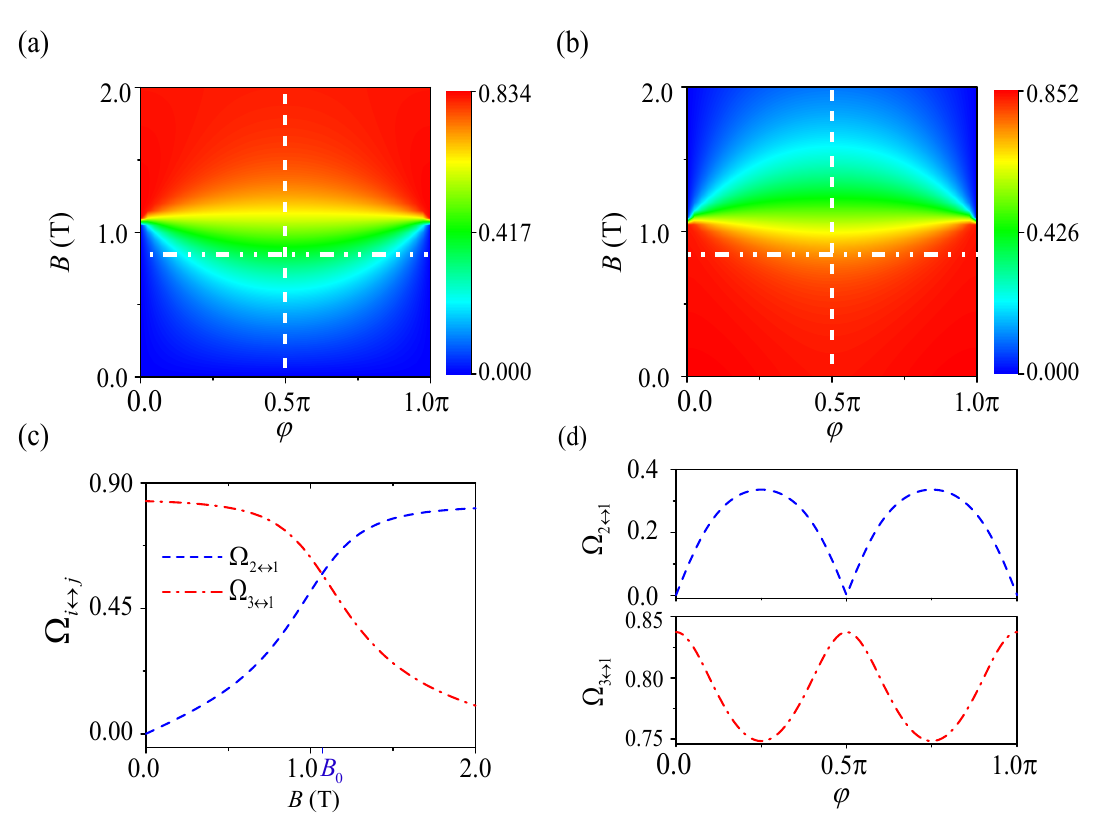}
\caption{(color online)(a) The electric-dipole transition rate $\Omega_{2\leftrightarrow 1}$ in units of $eEd/h$ as a function of the magnetic field strength $B$ and the angle $\varphi$. (b) The electric-dipole transition rate $\Omega_{3\leftrightarrow 1}$ in units of $eEd/h$ as a function of the magnetic field strength $B$ and the angle $\varphi$.  Panel (c) shows the variations of the transition rates with the magnetic field strength $B$, when $\varphi=\pi/2$; while panel (d) demonstrates the controllability of the transition rates by regulating the angle $\varphi$ when $B=0.8$ T.  The blue dashed curve represents $\Omega_{2\leftrightarrow 1}$, and the red dot-dashed curve corresponds to $\Omega_{3\leftrightarrow1}$. For a fixed SOC, the magnitude of $\varphi$ can be changed by varying the magnetic field direction.  The results are for an InSb nanowire DQD with the half interdot distance $d=40$nm.   }
\label{tranb}
\end{figure}

In Sec.2 we have shown that there are two mechanisms leading to different spin states hybridization in the eigenstates of DQD: the SOC-induced intradot states mixing and the interdot spin-flipped tunneling. Because all the mechanisms show strong dependences on the external magnetic field, both the transition rates $\Omega_{2\leftrightarrow1}$ and $\Omega_{3\leftrightarrow1}$ will definitely change when varying the magnetic field. As is clearly illustrated in Fig.\ref{tranb}, the variations of $\Omega_{2\leftrightarrow1}$ and $\Omega_{3\leftrightarrow1}$ with the magnetic field strength $B$ and orientation $\varphi$
are shown.

For an InSb nanowire DQD with $\xi_{\kappa}\eta_{\kappa} e^{-\eta_{\kappa}^{2}}\ll e^{-d^{2}/\bar{x}^{2}}\ll1$, the interdot tunneling dominates the orbital mixing in the eigenstates of DQD, and the effect of the intradot orbital states mixing can be negligible (the effect of the intradot orbital states mixing is investigated later in the next subsection).In a weak magnetic field, $B\ll B_{0}$, the major pseudospin components of the state $|\Phi_{1}\rangle$ are the same as that of $|\Phi_{3}\rangle$ and different from that of $|\Phi_{2}\rangle$.  It follows naturally that $\Omega_{3\leftrightarrow 1}\gg \Omega_{2\leftrightarrow 1}$. For a fixed $\varphi$ (the angle between the applied magnetic field and the SOC-induced effective field), increasing the magnetic field strength enhances the degree of the interdot pseudospin hybridization in $|\Phi_{2}\rangle$ and $|\Phi_{3}\rangle$, which in turn leads to the rising (falling) of $\Omega_{2\leftrightarrow 1}$ ($\Omega_{3\leftrightarrow 1}$), as shown in Fig.~\ref{tranb}(c).

$|\Phi_{1} \rangle \leftrightarrow |\Phi_{2}\rangle$ corresponds to the electric-dipole spin transition for $B<B_{0}$, with $\Omega_{2 \leftrightarrow 1}$ representing the EDSR frequency when the AC electric field is on resonance with $\Delta_{12}$. As demonstrated in Refs.~\citeonline{Rashba2003,Golovach2006}, the magnitude of the EDSR frequency depends on the effective SOC strength, which can be controlled by changing the magnetic field direction. In Fig.~\ref{tranb}(d), for a fixed magnetic field strength, the magnitude of the EDSR frequency $\Omega_{2 \leftrightarrow 1}$ as a function of the field orientation $\varphi$ is shown.  In particular, when the magnetic field is perpendicular to the SOC field direction, the effect of the SOC-induced mixing reaches its maximum, and the EDSR frequency reaches its peak value.  Similarly, $\Omega_{3 \leftrightarrow 1}$ also has a strong $\varphi$ dependence.

As $B$ increases beyond $B_0$, the major pseudospin components of $|\Phi_{2}\rangle$ and $|\Phi_{3}\rangle$ are swapped. At this point, $|\Phi_{1} \rangle \leftrightarrow |\Phi_{3} \rangle$ is the spin-flip transition, with $\Omega_{3 \leftrightarrow 1}$ the corresponding EDSR frequency.  In the increasing magnetic field, the larger energy splitting between $|\Phi_{2}\rangle$ and $|\Phi_{3}\rangle$ weakens the interdot pseudospin hybridization in these levels. As a result, the EDSR frequency $\Omega_{3 \leftrightarrow 1}$ decreases, and the orbital transition rate $\Omega_{2 \leftrightarrow 1}$ saturates, as shown in Fig.~\ref{tranb}(c).

\subsection{The effect of the intradot spin mixing }

In a DQD the interdot state mixing decreases exponentially with the increase of the interdot distance. When the interdot distance increases to a certain extend, the intradot orbit states mixing $\xi_{\kappa}\eta_{\kappa} e^{-\eta_{\kappa}^{2}}$ becomes comparable to the interdot overlaps $e^{-d^{2}/\bar{x}^{2}}$ in our considered range of magnetic field, so that intradot orbital mixing becomes an important factor in determining the overall spin-flip transition rates.  Here we examine the competition between the interdot and intradot mechanisms for spin flip transitions.

\begin{figure}[!ht]
\centering
\includegraphics[width=0.40\textwidth]{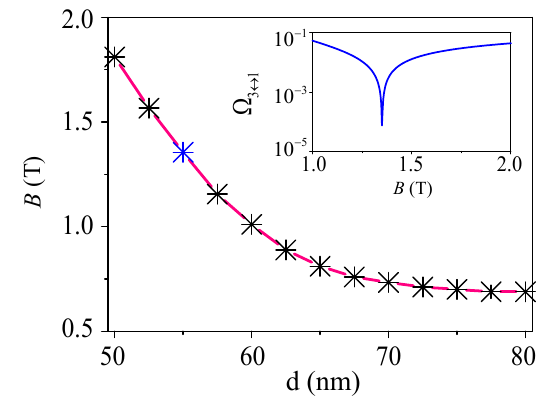}
\caption{(color online) The turning magnetic field $B_{\rm t}$ as a function of the half interdot distance $d$ when $\varphi=\pi/2$. The inset shows the electric-dipole transition rate $\Omega_{3\leftrightarrow1}$ (the EDSR frequency), in units of $eEd/h$, as a function of $B$ around the turning point $B_{t}$ when the half interdot distance $d=55$ nm.   }
\label{turn}
\end{figure}

In a low magnetic field, the effect of the intradot orbital states mixing on the EDSR, compared with the interdot mechanism, is negligible as long as $\xi_{\kappa}$ is small. As the Zeeman splitting increases, the rising value of $\xi_{\kappa}$ enhances the strength of the intradot orbital states mixing, see Eq.(\ref{mix}). Meanwhile, the interdot pseudospin states mixing weakens with the increase of $B$ for $B>B_{0}$.  There thus exists a turning magnetic field $B_{\rm t}$: for $B<B_{\rm t}$, EDSR is dominated by the interdot state hybridization; for high fields the state mixing is dominated by the intradot mechanism.  This change is also reflected in the variation of the EDSR frequency $\Omega_{3\leftrightarrow 1}$ around the turning field $B_{\rm t}$, as shown in the inset of Fig.~\ref{turn}. As the $B$ field increases and approaches $B_{\rm t}$, the magnitude of the EDSR frequency $\Omega_{3\leftrightarrow 1}$ decreases with the growth of $B$ as the interdot state mixing mechanism becomes less efficient, so it reverts that trend when $B>B_{\rm t}$ as the intradot mechanism becomes more effective.

The turning field $B_{\rm t}$ is a symbol for the competition between these two different mechanisms. Its magnitude mainly depends on the interdot distance. For the InSb nanowire DQD with $\varphi=\pi/2$, $B_{\rm t}$ as a function of $d$ is shown in Fig.~\ref{turn}. The downward trend of $B_{\rm t}$ with the increase of $d$ can be explained by the decline of the interdot state mixing, which requires a smaller magnetic field to counteract.

At large interdot distances, the magnitude of $B_{t}$ tends to be stable. This is because at a large interdot distance intradot orbital mixing dominates over interdot pseudospin hybridization, even for smaller magnetic field $B \leq B_{0}$.  Now $B_{0}$ mainly depends on the orbital energy difference between the QDs $\Delta_{o}\equiv\hbar\omega_{r}-\hbar\omega_{l}$, and nearly independent of $d$, $B_{0}\simeq \Delta_{o}/[(g_{el}+g_{er})\mu_{B}]$. With our chosen QD parameters, we find $B_{0}\simeq 0.688$ T.  Thus, once $B$ increases beyond $B_{0}$, the electric-dipole spin transition $|\Phi_{1} \rangle \leftrightarrow |\Phi_{3}\rangle$ is dominated by the intradot orbital mixing, and the EDSR frequency increases with $B$.

\subsection{The dependence on the interdot distance}

\begin{figure}[!ht]
\centering
\includegraphics[width=0.45\textwidth]{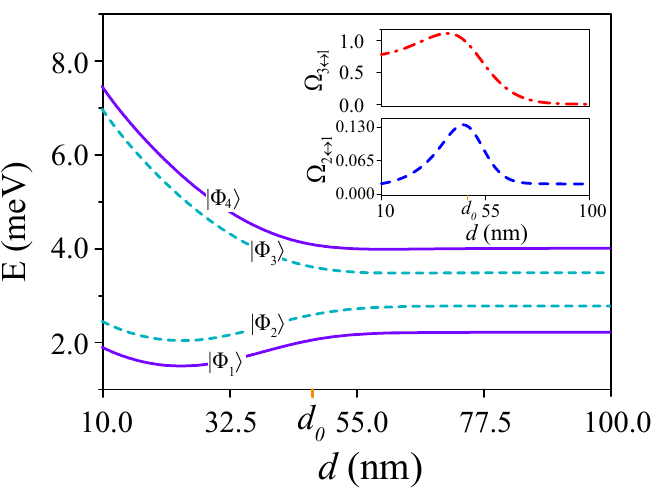}
\caption{(color online)   Energy spectrum of the nanowire DQD as a function of the half interdot distance $d$, at a magnetic field of $B=0.3$~T and an orientation of $\varphi=0.5\pi$. The inset shows the electric-dipole transition rates $\Omega_{2\leftrightarrow 1}$ and $\Omega_{3\leftrightarrow 1}$, in units of $eE\bar{x}/h$, as functions of $d$.}
\label{enerd}
\end{figure}

The underlying dependence of the interdot barrier on the interdot distance means that spin tunneling, and single-electron energy spectrum of the nanowire DQD in general, depend on $d$~\cite{Jr2011}.  In Fig.~\ref{enerd} we show the energy spectrum of the nanowire InSb DQD as a function of $d$, with $B=0.3$~T and $\varphi=\pi/2$.
Because of the asymmetry in the confinement potential along the wire axis, we limit ourselves to consider the case with a nonzero finite interdot distance exclusively.

At a large interdot distance, the interdot tunneling (proportional to $e^{-d^{2}/\bar{x}^{2}}$) is much smaller than the orbital energy difference $\Delta_{o}$ between the dots, so that the effect of the interdot states mixing on the energy spectrum is negligible.  The energy spectrum of the DQD is essentially the sum of the energy spectrums of the single QDs in this case, with $|\Phi_{1}\rangle\simeq |\Psi^{+}_{l}\rangle$, $|\Phi_{2}\rangle\simeq |\Psi^{-}_{l}\rangle$, $|\Phi_{3}\rangle\simeq |\Psi^{+}_{r}\rangle$ and $|\Phi_{4}\rangle\simeq |\Psi^{-}_{r}\rangle$. When the interdot distance decreases, the interdot tunnel coupling increases exponentially, and the energy spectrum of the DQD changes correspondingly. When $d \sim d_{0}$, the energy scale of the interdot tunnel coupling becomes comparable to the orbital energy difference between the QDs, so that the eigenstates of the DQD are delocalized pseudospin states.  If the interdot distance further decreases, the two dots start to merge.  The ``interdot tunneling'' will be of the same magnitude as the orbital excitation energy in the individual QDs.  At this limit, the character of the electronic states shifts back from molecular-like to atomic-like again like the case of large interdot distance~\cite{Jr2011}, although the composition of the orbital states are dramatically different.  The energy splitting between $|\Phi_{2}\rangle$ and $|\Phi_{3}\rangle$ is now dominated by the single-particle single-dot excitation energy (as compared to tunnel splitting in the case of a double dot), which results in a sharp rise in this energy gap, as shown in Fig.~\ref{enerd}.

Since the compositions of the DQD eigenstates vary with the interdot distance, particularly near $d_{0}$, the electric-dipole transition rates change quite dramatically as well.  In the inset of Fig.~\ref{enerd} we plot the transition rates $\Omega_{2\leftrightarrow 1}$ and $\Omega_{3\leftrightarrow 1}$ as a function of $d$. At a large interdot distance $d\gg d_{0}$, the DQD eigenstates can be approximated as the eigenstates of the individual QDs, as explained above. Thus, $|\Phi_{2}\rangle\leftrightarrow |\Phi_{1}\rangle$ is an intradot spin-flip transition while $|\Phi_{3}\rangle \leftrightarrow |\Phi_{1}\rangle$ is an interdot transition. Because of the vanishingly small interdot state mixing, the magnitude of the interdot transition rate will be smaller than that of the intradot spin-flip transition rate, $\Omega_{3\leftrightarrow 1}<\Omega_{2\leftrightarrow 1}$. As the interdot distance decreases, the rapidly rising interdot coupling means both transition rates increase quickly as the states become mixed. When $d \sim d_{0}$, the eigenstates of the DQD are delocalized, and the electric-dipole spin transition $|\Phi_{2}\rangle\leftrightarrow |\Phi_{1}\rangle$ is dominated by the interdot pseudospin tunneling.  As $d$ decreases further, the magnitudes of $\Omega_{2\leftrightarrow1}$ and $\Omega_{3\leftrightarrow1}$ become stable because the DQD merges into a single QD.  The electric-dipole transition in a single nanowire QD was investigated in Ref.~\citeonline{Ruili2013}. In this limit, $\Omega_{3 \leftrightarrow 1}$ approaches $(\sqrt{2}/2)eE\bar{x}/h$, while $\Omega_{2 \leftrightarrow 1}$ can be approximated by $\xi_{l}\eta_{l} \exp(-\eta_{l}^{2})eEx_{l}/h$. Thus, when $d\ll d_{0}$ the main mechanism of the EDSR turns back to the intradot orbital states mixing again. In short, in the parameter range we have considered, the electric-dipole transition rates depend sensitively on the interdot tunneling/distance.

\section{ Phonon-induced relaxation between the energy levels}

Electron-phonon (e-ph) interaction, together with spin-orbit coupling, is the main cause of spin relaxation in a quantum dot~\cite{Khaetskii2001,Khaetskii2000, Bulaev2005, Florescu2006, Golovach2004, Fabian20061, Fabian20062, Scarlino2014,Hanson2003,Romano2008,Vladimir2005,Raith2012,Destefani2005}. Accurately determining the relaxation rates is thus a necessary condition for quantitatively assessing the fidelity of the electric-dipole transitions. Recall that $B_0$ is the field at which $|\Phi_{2}\rangle$ and $|\Phi_{3}\rangle$ cross in the absence of SOC.  Based on the major pseudospin components of the eigenstates involved in relaxation, $\Gamma_{2\rightarrow 1}$ corresponds to phonon-induced spin relaxation for $B \ll B_{0}$, while $\Gamma_{3\rightarrow 1}$ is the phonon-induced spin relaxation rate when $B \gg B_{0}$.

For relaxation between energetically close levels, we only consider the e-ph interaction with acoustic phonons and ignore the optical phonons. For acoustic phonons, there are two types of e-ph interaction: the piezoelectric and deformation potential interactions~\cite{Cleland}. Including the e-ph interaction, the complete Hamiltonian describing the DQD reads
\begin{align}
H_{\rm tot}=H_{\rm DQD}+V_{\rm e-ph},
\end{align}
where the e-ph interaction is given by~\cite{Fabian20061,Fabian20062}
\begin{align}
V_{\rm e-ph}=&\sum_{q,\lambda}\sqrt{\frac{\hbar q}{2\rho V c_{\lambda}}}(V^{\rm df}_{\mathbf{q},\lambda}-iV^{\rm pe}_{\mathbf{q},\lambda})(b_{\mathbf{q},\lambda}+b^{\dagger}_{-\mathbf{q},\lambda})e^{i\mathbf{q}\cdot\mathbf{r}} \,.
\label{e-ph}
\end{align}
For the deformation potential interaction, $V^{\rm df}_{\mathbf{q},\lambda}=D_{e}\delta_{\lambda,l}$; and for the piezoelectric interaction, $V^{\rm pe}_{\mathbf{q},\lambda} = 2eh_{14} (q_{x}q_{y}\hat{e}^{\lambda}_{\mathbf{q},z} + q_{y}q_{z} \hat{e}^{\lambda}_{\mathbf{q},x} + q_{z}q_{x} \hat{e}^{\lambda}_{\mathbf{q},y})/q^{3}$.  Here $\mathbf{q}=(q_{x},~q_{y},~q_{z})$ is the phonon wave vector, with $q$ representing its magnitude, $\mathbf{r}=(x,~y,~z)$ denotes the electron position, and $\lambda$ is the polarization of the phonon, with $\hat{e}$ and $c_{\lambda}$ being the polarization vector and sound velocity of the phonon mode.  The phonon annihilation (creation) operator is denoted by $b$ ($b^{\dagger}$).  $\rho$ and $V$ are the mass density and the volume of the sample, respectively.

\begin{figure}[!ht]
\centering
\includegraphics[width=0.42\textwidth]{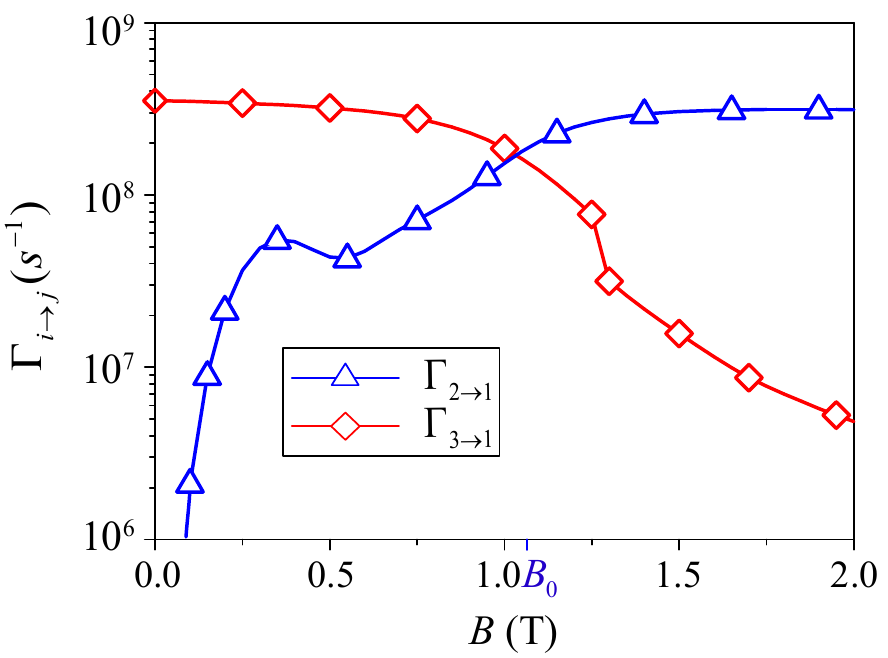}
\caption{(color online) The phonon-induced relaxation rates $\Gamma_{i\rightarrow j}$ as a function of the magnetic field strength $B$ with $\varphi=\pi/2$ and $d=40~\rm nm$.}
\label{relaxation}
\end{figure}

Using the Fermi golden rule, the phonon-induced relaxation rate between the eigenstates $|\Phi_{i}\rangle$ and $|\Phi_{j}\rangle$ ($i>j$) can be calculated as
\begin{align}
\Gamma_{i\rightarrow j}=&\sum_{q,\lambda}\frac{\pi q}{\rho V c_{\lambda}}|V^{\rm df}_{\mathbf{q},\lambda}-iV^{\rm pe}_{\mathbf{q},\lambda}|^{2}|M_{ij}|^{2} [n_{\rm th}(T)+1]\delta(\hbar c_{\lambda}q-\Delta_{ij}),
\end{align}
where $\Delta_{ij}$ is the energy difference between the eigenstates $|\Phi_{i}\rangle$ and $|\Phi_{j}\rangle$, $n_{\rm th}$ is the thermal occupation of the phonon mode with $\hbar\omega_{\lambda}=\Delta_{ij}\equiv E_{i}-E_{j}$.  At low temperatures when $k_{\rm B}T \ll \Delta_{ij}$, $n_{\rm th} \approx 0$. The matrix element $M_{ij}$ depends on the spatial distribution of the electron wave functions in three dimensions (see Methods).  Taking the phonon mode density into consideration, the phonon-induced relaxation rate can be rewritten as
\begin{align}
\Gamma_{ i\rightarrow j}=\sum_{\lambda}\frac{1}{8\hbar\pi^{2}\rho c^{2}_{\lambda}}\oint q |(V^{\rm df}_{\mathbf{q},\lambda}-iV^{\rm pe}_{\mathbf{q},\lambda})M_{ij}|^{2} d \mathbf{q},
\label{relaxations}
\end{align}
with the integral region satisfying the energy conservation condition $\hbar c_{\lambda}q = \Delta_{ij}$.

The electron-phonon interaction Hamiltonian in Eq.(\ref{e-ph}) is spin independent. As such the spin composition of $|\Phi_{i}\rangle$ and $|\Phi_{j}\rangle$ plays a key role in determining $\Gamma_{i\rightarrow j}$: a mostly spin-conserved relaxation would be much faster than a relaxation involving spin flip.  Since the degree of spin mixing depends on the interplay between the external field and the spin-orbit coupling, the relaxations between different eigenstates can generally be regulated by varying the magnetic field strength and direction~\cite{Fabian20061,Fabian20062, Golovach2004, Scarlino2014, Romano2008,Raith2012,Vladimir2005,Destefani2005}.  Here we focus on the dependence of the relaxation rates on the magnetic field strength for a fixed magnetic field direction. The numerical results are shown in Fig.~\ref{relaxation}, where we plot $\Gamma_{2\rightarrow 1}$ and $\Gamma_{3\rightarrow 1}$ as functions of the magnetic field strength, with the system parameters taking the values as in Fig.\ref{enerb}.

In a low magnetic field with $B\ll B_{0}$, the  major pseudospin components of $|\Phi_{i}\rangle$ ($i=1-4$) dictate that $|\Phi_{3}\rangle\rightarrow |\Phi_{1}\rangle$  is a charge transition, while $|\Phi_{2}\rangle\rightarrow |\Phi_{1}\rangle$ is a spin-flip transition, thus $\Gamma_{3 \rightarrow 1} \gg \Gamma_{2 \rightarrow 1}$.  As $B$ increases, spin-flipped tunneling results in the hybridization between $|\Phi^{0}_{2}\rangle$ and $|\Phi^{0}_{3}\rangle$, so that $\Gamma_{3 \rightarrow 1}$ decreases while $\Gamma_{2 \rightarrow 1}$ increases.  The slight oscillations in both relaxation rates are most likely due to the matching between DQD charge density and phonon wave vector (recall that the e-ph interaction Hamiltonian contains a $e^{i\mathbf{q}\cdot\mathbf{r}}$ factor).  Furthermore, with the large $g$-factor for InSb, Zeeman splitting reaches 1 meV when the magnetic field is only a fraction of 1 Tesla.  The corresponding phonon wave length is in the order of 10 nm, already below the quantum dot size, so that phonon bottleneck effect starts to become apparent for spin-flip relaxation~\cite{Stroscio1992}.

When $B = B_{0}$ the spin states of $|\Phi_{2}\rangle$ and $|\Phi_{3}\rangle$ are equally mixed.  The energy gap between $|\Phi_{2} \rangle$ and $|\Phi_{3} \rangle$ means that the relaxation rates are generally not identical at $B_0$.  The rates are determined by a competition mostly between phonon density of states consideration and the phonon bottleneck effect: the former favors the larger-energy $3 \rightarrow 1$ transition, while the latter favors the smaller-energy $2 \rightarrow 1$ transition.

When the magnetic field strength exceeds $B_{0}$, $|\Phi_{2}\rangle$ ($|\Phi_{3}\rangle$) becomes the pseudospin up (down) state.  As illustrated in Fig.~\ref{enerb}, for $B > B_{0}$ the energy splitting $\Delta_{21}$ (proportional to the tunnel coupling in the DQD) tends to be stable with the growth of $B$, while $\Delta_{31}$, now the Zeeman splitting, keeps increasing. Thus $\Gamma_{2\rightarrow 1}$ approaches a constant value when $B \gg B_{0}$. $\Gamma_{3\rightarrow1}$, on the other hand, keeps decreasing due to the reduction in spin mixing and the increasing influence of the phonon bottleneck effect.

The relaxation calculation here is done using bulk phonons.  In a suspended nanowire, confined phonons on the nanowire should be used, and we expect the relaxation rates to be further suppressed because of the much smaller phonon density of states and stronger anisotropy due to the nanowire geometry~\cite{Nishigushi1994,Loss2008}.

\section*{ Discussion and Conclusion}

In this paper, we study the electronic properties of a nanowire DQD within the frame of effective mass approximation (EMA). For a thin nanowire, the energy scale of the electronic dynamics along the axis direction is much smaller than the energy scale of the excitations in the transverse directions.  As such in our consideration the confined electron always stays in the transverse ground state~\cite{Osika2013, Nowak2013, Loss2008, Deng2016}.

Our calculations are based on a truncated double harmonic potential.  Within EMA, the DQD confinement potential is usually approximated by a quartic function, a biquadratic function, or a Gaussian function~\cite{Hu2006,Stano2005, Hu2000, Yang2011, Jr2011}. All these model potentials give rise to results consistent with the experimental results at a qualitative level~\cite{Hu2006,Yang2011,Fabian20061, Raith2012}.  Therefore, the simplicity associated with the truncated double harmonic well (biquadratic function) model potential becomes the deciding factor for our choice.  The relatively concise expressions within this model allows us to get to the basic physics more easily.

The system parameters used in our calculations are taken from the experimental data in Ref.~\citeonline{Perge2012}. The low-energy spectrum of a single electron in the DQD is obtained using the LCAO method.  In the calculation, the SOC is taken into account precisely, while the applied magnetic field is treated as a perturbation.

Our calculations show that in a DQD, there exist two different mechanisms that lead to EDSR: the intradot pseudospin state mixing and the interdot spin-flipped tunneling. The EDSR frequency is determined by the combined effect of these two mechanisms, in which the dominant role can be varied by changing the system parameters.  When the EDSR is dominated by the interdot spin-flipped tunneling, we show that the electric-dipole transition rates depend sensitively on the magnitude and orientation of the applied field.
The intradot orbital mixing becomes more important when we reduce the tunnel coupling, so the two dots become independent, or when we increase tunneling to the degree when the double dot merges into a single dot.  In the intermediate regime the interdot spin mixing is more effective.  For a fixed tunnel coupling/interdot distance, the electric-dipole driven transition rates experience a dip as the magnetic field increases, when the DQD transitions from the interdot-mixing dominated low-field region to the intradot-mixing dominated high-field region.

Finally, we have calculated phonon-induced relaxation rates among the DQD energy levels.  The very large $g$-factors for strong SOC materials, such as InSb that we consider, mean that phonon bottleneck effect kicks in at much lower magnetic field for spin-flip transitions compared to materials such as GaAs. Overall, our results on low-energy spectrum, controllable electric-dipole transitions, and relaxations should provide useful input for experimental studies of quantum coherent manipulations in a nanowire DQD.

\section*{Methods}

\textbf{Derivation of the effective Hamiltonian.}
The nanowire DQD Hamiltonian in Eq.~(\ref{Ham}) is derived within the effective mass approximation.  We choose our coordinate system according to the geometry of the nanowire and the applied field.  Specifically, we choose the $x$-axis along the axis of the nanowire, as illustrated in Fig.~\ref{model}(a).  When an external magnetic field is applied, we choose it to lie in the $xz$-plane, so that the field can be expressed as $\mathbf{B}=B(\cos\theta,0,\sin\theta)$.  The complete Hamiltonian describing an electron in a nanowire DQD is
\begin{align}
H(\mathbf{r})=\frac{\mathbf{P}^{2}}{2m_{e}}+U(\mathbf{r})+H_{\rm so}(\mathbf{r})+\frac{g_{e}\mu_{B}\mathbf{B}}{2}\cdot\boldsymbol{\sigma},
\label{ora}
\end{align}
where the first term is the kinetic energy, with the kinetic momentum $\mathbf{P}=\mathbf{p}+e\mathbf{A}$ and the vector potential $\mathbf{A} = B(-y \sin\theta,-z \cos\theta,0)$, $U(\mathbf{r})$ is the confinement potential in three dimensions, $H_{\rm so}(\mathbf{r})$ represents the spin-orbit interaction, and the last term denotes the Zeeman term, with $g_{e}$ and $\mu_{B}$ being the location-dependent Land\'{e} $g$-factor and Bohr magneton, respectively. Here $g_{e}$ is location-dependent, with the specific value of the $g$-factor of the left QD being different from that of the right QD $g_{el}\neq g_{er}$ .

The DQD confinement potential for the electron is modelled by a asymmetric double well harmonic potential along the nanowire axis, $V(x)=\frac{1}{2}m_{e}\min\{\omega_{l}^{2}(x+d)^{2},\omega_{r}^{2}(x-d)^{2}\}$, where $\omega_{l}\neq\omega_{l} $, and $2d$ is the interdot distance.  In the transverse direction we consider a strong harmonic potential along the $y$ direction, $V(y) = (1/2) m_{e} \omega^{2}_{y} y^{2}$, and a large gradient potential along the $z$ direction, $V(z)=eE_{z}z$ for $z \geq 0$ and $V(z) = \infty$ for $z < 0$. Due to the strong transverse confinements, we assume that the electron is always in the ground state along the $y$ and $z$, so that the transverse orbital dynamics is frozen:
\begin{align}
\psi(y)=&\frac{1}{\sqrt{\pi^{1/2}y_{0}}}\exp\left(-\frac{y^{2}}{2y^{2}_{0}}\right), \nonumber \\
\phi(z)=&  1.4261\sqrt{\tau}Ai(\tau z-2.3381),
\label{wavefunctions}
\end{align}
where $y_{0}=\sqrt{\hbar/(m_{e}\omega_{y})}$ and $\tau = (2eE_{z} m_{e}/\hbar^{2})^{1/3}$. The characteristic length scales of the wavefunction along the $y$ and $z$ directions can thus be quantified by $y_{0}$ and $z_{0}\equiv \int^{\infty}_{0}\phi(z)z\phi(z) dz = 1.5581/\tau$, respectively.

The lowest-order effective Hamiltonian for an electron moving along the $x$-axis can be obtained by averaging over the $y$ and $z$ directions,
\begin{align}
H_{\rm DQD}\equiv&\left\langle H(\mathbf{r})\right\rangle \nonumber\\
=&\left\langle \frac{\mathbf{P}^{2}}{2m_{e}}\right\rangle+\left\langle H_{\rm so}(\mathbf{r})\right\rangle+V(x)+\frac{g_{e}\mu_{B}\mathbf{B}}{2}\cdot\boldsymbol{\sigma},
\label{xham}
\end{align}
where $\langle\xi\rangle=\int\psi^{\ast}(y)\phi^{\ast}(z)\xi\psi(y)\phi(z)dydz$.  The first term on the right side of Eq.(\ref{xham}) represents the effective kinetic Hamiltonian. Substituting the kinetic momentum expression into Eq.~(\ref{xham}), the effective kinetic Hamiltonian can be expanded as
\begin{align}
\left\langle \frac{\mathbf{P}^{2}}{2m_{e}}\right\rangle=& \int \psi^{\ast}(y) \phi^{\ast}(z) \left[\frac{(p_{x}-eB_{z}y)^{2}}{2m_{e}} + \frac{(p_{y}-eB_{x}z)^{2}}{2m_{e}} + \frac{p_{z}^{2}}{2m_{e}}\right] \psi(y) \phi(z) dydz \nonumber\\
=&\frac{p^{2}_{x}}{2m_{e}}-eB_{z}\frac{p_{x}}{m_{e}}\langle y\rangle + \left[\frac{(eB_{z})^{2}\langle y^{2} \rangle}{2m_{e}} + \frac{\langle p_{z}^{2} \rangle}{2m_{e}} + \frac{\langle(p_{y}-eB_{x}z)^{2}\rangle}{2m_{e}} \right],
\label{fterm}
\end{align}
with $B_{x}=B\cos\theta$ and $B_{z}=B\sin\theta$. Since $\psi(y)$ is an even function, the second term on the right side of Eq.~(\ref{fterm}) vanishes. The last term on the right side of Eq.~(\ref{fterm}) can also be ignored because it is a constant term and only affects the zero-point energy of the effective Hamiltonian.  In short, the applied magnetic field does not have any orbital effect within this mean field approximation.

The inversion asymmetry in $\rm{A_{III}B_{V}}$ heterostructures results in Dresselhaus and Rashba spin-orbit interactions~\cite{Winkler2003,Sousa2003},
\begin{align}
H_{\rm so}(\mathbf{r}) = & H_{D}(\mathbf{r})+ H_{R}(\mathbf{r}), \nonumber \\
H_{D}(\mathbf{r}) = & \frac{\gamma_{D}}{2\hbar^{3}}\boldsymbol{\sigma}\cdot \mathbf{\tilde{P}}, \nonumber\\
H_{R}(\mathbf{r}) = &\frac{\gamma_{R}}{\hbar}\boldsymbol{\sigma}\cdot\nabla U(\mathbf{r})\times \mathbf{P}.
\label{specific SOC}
\end{align}
Here interaction strength $\gamma_{D}$ and $\gamma_{R}$ are determined by the band structure parameters~\cite{Winkler2003,Sousa2003}. $\tilde{P}_{x} = P_{x}(P^{2}_{y} - P^{2}_{z}) + \rm{H.c.}$, while $\tilde{P}_{y}$ and $\tilde{P}_{z}$ can be obtained by cyclic permutations.
The effective SOC Hamiltonian along the $x$ direction can thus be calculated as
\begin{align}
H^{x}_{\rm so}\equiv &\langle H_{\rm so}(\mathbf{r})\rangle, \nonumber \\
=& \langle H_{D}(\mathbf{r})\rangle + \langle H_{R}(\mathbf{r})\rangle.
\end{align}

According to Eq.~(\ref{specific SOC}), the effective Hamiltonian describing the linear Dresselhaus SOC along the nanowire axis is
\begin{align}
H^{x}_{D} \equiv &\frac{\gamma_{D}}{\hbar^{3}} \sigma_{x}p_{x} \left( \langle P^{2}_{y} \rangle-\langle P^{2}_{z} \rangle \right) ,\nonumber\\
= & \frac{\gamma_{D}}{\hbar^{3}}\sigma_{x}p_{x} \left[ \langle p^{2}_{y}\rangle-\langle p^{2}_{z}\rangle-(eB_{x})^{2}\langle z^{2}\rangle \right],
\label{Dresselhaus SOC}
\end{align}
where we have used the identity $\langle p_{y} \rangle=0$.  In the considered range of magnetic field with $\xi \ll 1$,  the contribution of $(eB_{x})^{2}\langle z^{2} \rangle$ to $H^{x}_{D}$ is negligible compared with the other two terms in the bracket on the right side of Eq.~(\ref{Dresselhaus SOC}).  Similarly, the effective Rashba SOC along the $x$ direction can be written as
\begin{align}
H^{x}_{R}\equiv &\langle H_{R}(\mathbf{r})\rangle ,\nonumber\\
=&\frac{\gamma_{R}}{\hbar}p_{x}\left(\sigma_{y}\langle\partial_{z}V(z)\rangle-\sigma_{z}\langle\partial_{y}V(y)\rangle\right).
\end{align}
Using the specific confinements along lateral directions, we obtain $\partial_{y} V(y) = m_{e}\omega^{2}_{y}y$ and $\partial_{z} V(z) = eE_{z}$. After averaging over $y$ and $z$, the effective Rashba SOC Hamiltonian takes the form
\begin{align}
H^{x}_{R}=\frac{\gamma_{R}}{\hbar}eE_{z}\sigma_{y}p_{x}.
\end{align}

The total effective SOC Hamiltonian along the nanowire axis is thus given by
\begin{align}
H^{x}_{\rm so} \equiv &\left\langle H_{\rm so}(\mathbf{r})\right\rangle \nonumber \\
=&\alpha_{R}\sigma_{y}p_{x}+\alpha_{D}\sigma_{x}p_{x},
\label{xsoc}
\end{align}
with
\begin{align}
\alpha_{ R}=\frac{\gamma_{R}}{\hbar}eE_{z},~\alpha_{ D}=\frac{\gamma_{D}}{\hbar^{3}}(\langle p^{2}_{y}\rangle-\langle p^{2}_{z}\rangle).
\end{align}

Substituting Eq.~(\ref{fterm}) and (\ref{xsoc}) into Eq.~(\ref{xham}), the effective Hamiltonian describing the DQD along the wire axis can be simplified as Eq.~(\ref{Ham}) in the main text,
\begin{align}
H_{\rm DQD}=\frac{p_{x}^{2}}{2m_{e}}+V(x)+\alpha p_{x}\sigma^{a}-\frac{\Delta_{Z}}{2}\sigma^{n},
\end{align}
with $\alpha=\sqrt{\alpha^{2}_{ R}+\alpha^{2}_{ D}}$ and $\Delta_{Z} = -g_{e}\mu_{B}B$.

In the numerical calculations in this paper, we assume $\hbar\omega_{y}=80$ meV and $E_{z}=0.6$ mV/{\AA}.  The SOC length in an InSb nanowire DQD is $x_{\rm so} = 200$ nm, and the characteristic lengths along the transverse directions are given by $y_{0}=8.2$ nm and $z_{0}=12.4$ nm. Other material parameters are all chosen for a nominal InSb nanowire, including $m_{e}=0.013m_{0}$, $\rho=5.77\times 10^{-27}~\rm kg/{\AA}^{3}$, $\gamma_{D}=228~\rm eV{\AA}^{3}$, $\gamma_{R}=500~\rm {\AA}^{2}$, $D_{e}=7.0$ eV, $eh_{14}=0.061$ eV/{\AA}, $c_{l}=3.69\times 10^{13}$ {\AA}/s, and $c_{t}=2.3\times 10^{13}$ {\AA}/s, which are used in the main text for numerical calculations.

\bigskip
\noindent \textbf{Construction of the orthonormal basis.}
The analytic formulas for the orthonormal bases $|\Psi_{\kappa\Uparrow}\rangle$ and $|\Psi_{\kappa\Downarrow}\rangle$ ($\kappa=l,r$) are given.
Using the perturbation theory, the two lowest-energy eigenstates of the local Hamiltonian $H_{\kappa}$ can be approximated as the equation (\ref{mix}) in the main text
\begin{align}
|\Psi^{\pm}_{\kappa}\rangle=&c^{\pm}_{\kappa}|\Phi_{\kappa 0\uparrow}\rangle+d^{\pm}_{\kappa}|\Phi_{ \kappa 0\downarrow}\rangle+i(\xi/2)e^{-\eta^{2}}\sin\varphi   \sum^{\infty}_{n=1}\frac{(\sqrt{2}\eta)^{n}}{n\sqrt{n!}}[(-i)^{n}c^{\pm}_{\kappa}|\Phi_{\kappa n \downarrow}\rangle-d^{\pm}_{\kappa}|\Phi_{\kappa n \uparrow}\rangle],
\label{apmix}
\end{align}
with the parameters given in Eqs.(\ref{mix2}). As is indicated by Eq.(\ref{mix}), the Zeeman field leads to the mixing of different spin-orbit states in $|\Psi^{\pm}_{\kappa}\rangle$, with the degree of orbital mixing proportional to $\xi_{\kappa}\eta_{\kappa}^{n}e^{-\eta_{\kappa}^{2}}$. Here  $\xi_{\kappa}$ denotes the ratio between the Zeeman splitting and the orbital splitting in $\kappa$ dot, $\xi_{\kappa}\equiv\Delta_{\kappa Z}/\Delta_{\kappa S}$, and is much less than one, which ensures the validity of the perturbation theory. $\eta_{\kappa}$ corresponds to the ratio between the effective dot size $x_{\kappa}$ and SOC length $x_{\rm so}$. In a nanowire quantum dot,  $\eta_{\kappa}$ is generally a small number $\eta_{\kappa} \equiv x_{\kappa}/x_{\rm so}\ll 1$, even for materials with strong SOC. Therefore, in order to facilitate the numerical calculations in the main text and account the effect of high orbital states, the summation in Eq.(\ref{apmix}) is truncated, and only keep the $n=1$ term.
Thus, the corresponding normalized local wave functions can be written as
\begin{align}
|\Psi^{+}_{1l}\rangle=&\frac{\cos\vartheta_{l}}{\sqrt{\chi^{2}_{l}+1}}(|\Phi_{l 0\uparrow }\rangle+\chi_{l}|\Phi_{l 1\downarrow }\rangle)
-i\frac{\sin\vartheta_{l}}{\sqrt{\chi^{2}_{l}+1}}(|\Phi_{l 0\downarrow }\rangle+\chi_{l}|\Phi_{l 1\uparrow }\rangle), \nonumber \\
|\Psi^{-}_{1l}\rangle=&\frac{\sin\vartheta_{l}}{\sqrt{\chi^{2}_{l}+1}}(|\Phi_{l 0\uparrow }\rangle+\chi_{l}|\Phi_{l 1\downarrow }\rangle)
+i\frac{\cos\vartheta_{l}}{\sqrt{\chi^{2}_{l}+1}}(|\Phi_{l 0\downarrow }\rangle+\chi_{l}|\Phi_{l 1\uparrow }\rangle), \nonumber \\
|\Psi^{+}_{1r}\rangle=&\frac{\cos\vartheta_{r}}{\sqrt{\chi^{2}_{r}+1}}(|\Phi_{r 0\uparrow }\rangle+\chi_{r}|\Phi_{r 1\downarrow }\rangle)
-i\frac{\sin\vartheta_{r}}{\sqrt{\chi^{2}_{r}+1}}(|\Phi_{r 0\downarrow }\rangle+\chi_{r}|\Phi_{r 1\uparrow }\rangle), \label{speceqs} \\
|\Psi^{+}_{1r}\rangle=&\frac{\sin\vartheta_{r}}{\sqrt{\lambda^{2}_{r}+1}}(|\Phi_{r 0\uparrow }\rangle+\chi_{r}|\Phi_{r 1\downarrow }\rangle)
+i\frac{\cos\vartheta_{r}}{\sqrt{\chi^{2}_{r}+1}}(|\Phi_{r 0\downarrow }\rangle+\chi_{r}|\Phi_{r 1\uparrow }\rangle),\nonumber
\end{align}
with $\chi_{\kappa}=(\sqrt{2}/2)\xi_{\kappa}\eta_{\kappa} e^{-\eta_{\kappa}^{2}}\sin\varphi$ and $\vartheta_{\kappa}=(1/2)\arccos(\cos\varphi/f_{\kappa})$.

On the basis of Eqs.(\ref{speceqs}), we can construct the two orthonormal bases
\begin{align}
|\Psi_{l\Downarrow}\rangle=\frac{1}{\sqrt{1+|g_{-}|^{2}-2{\rm {Re} }[g_{-}s_{-}]}}(|\Psi^{-}_{1l}\rangle-g_{-}|\Psi^{-}_{1r}\rangle), \nonumber\\
|\Psi_{r\Downarrow}\rangle=\frac{1}{\sqrt{1+|g_{-}|^{2}-2{\rm {Re} }[g_{-}s_{-}]}}(|\Psi^{-}_{1r}\rangle-g^{\ast}_{-}|\Psi^{-}_{1l}\rangle)
\end{align}
where $s_{-}=\langle\Psi^{-}_{1l}|\Psi^{-}_{1r}\rangle$ and $g_{-}=(1-\sqrt{1-|s_{-}|^{2}})/s_{-}$. In order to construct the other two orthonormal bases, first we introduce two auxiliary states
\begin{align}
|\widehat{\Psi}_{l\Uparrow}\rangle=\frac{1}{\sqrt{1-|s_{1l}|^{2}-|s_{2l}|^{2}}}(|\Psi^{+}_{1l}\rangle-s_{1l}|\Psi_{l\Downarrow}\rangle-s_{2l}|\Psi_{r\Downarrow}\rangle), \nonumber\\
|\widehat{\Psi}_{r\Uparrow}\rangle=\frac{1}{\sqrt{1-|s_{1r}|^{2}-|s_{2r}|^{2}}}(|\Psi^{+}_{1r}\rangle-s_{1r}|\Psi_{l\Downarrow}\rangle-s_{2r}|\Psi_{r\Downarrow}\rangle)
\end{align}
with $s_{1l}=\langle\Psi^{-}_{l\Downarrow}|\Psi^{+}_{1l}\rangle$, $s_{2l}=\langle\Psi^{-}_{r\Downarrow}|\Psi^{+}_{1l}\rangle$, $s_{1r}=\langle\Psi^{-}_{l\Downarrow}|\Psi^{+}_{1r}\rangle$, and $s_{2r}=\langle\Psi^{-}_{r\Downarrow}|\Psi^{+}_{1r}\rangle$. Finally, basing on the auxiliary states, the other two orthonormal bases can be calculated as
\begin{align}
|\Psi_{l\Uparrow}\rangle=\frac{1}{\sqrt{1+|g_{+}|^{2}-2{\rm {Re} }[g_{+}s_{+}]}}(|\widehat{\Psi}_{l\Uparrow}\rangle-g_{+}|\widehat{\Psi}_{r\Uparrow}\rangle), \nonumber\\
|\Psi_{r\Uparrow}\rangle=\frac{1}{\sqrt{1+|g_{+}|^{2}-2{\rm {Re} }[g_{+}s_{+}]}}(|\widehat{\Psi}_{r\Uparrow}\rangle-g^{\ast}_{+}|\widehat{\Psi}_{l\Uparrow}\rangle),
\end{align}
where $s_{+}=\langle\widehat{\Psi}_{l\Uparrow}|\widehat{\Psi}_{r\Uparrow}\rangle$ and $g_{+}=(1-\sqrt{1-|s_{+}|^{2}})/s_{+}$.

\bigskip
\noindent\textbf{Calculation of the phonon-induced relaxation rates.}
For relaxation between energy levels of a nanowire DQD through a single-phonon process, we only consider the e-ph interaction with acoustic phonons and ignore the optical phonons. For acoustic phonons, there are two types of e-ph interaction: the piezoelectric and deformation potential e-ph interactions~\cite{Cleland}. The corresponding Hamiltonian is given by Eq.(\ref{e-ph}).

At low temperatures with $k_{\rm B}T \ll \Delta_{ij}$, the phonon-induced relaxation rate between states $|\Phi_{i}\rangle$ and $|\Phi_{j}\rangle$ ($i>j$) can be calculated via the Fermi golden rule:
\begin{align}
\Gamma_{ i\rightarrow j}=&\sum_{q,\lambda}\frac{\pi q}{\hbar\rho V c_{\lambda}}|V^{\rm df}_{\mathbf{q},\lambda}-iV^{\rm pe}_{\mathbf{q},\lambda}|^{2}|M_{ij}|^{2} \delta(\hbar c_{\lambda}q-\Delta_{ij}),
\end{align}
where $\Delta_{ij}$ denotes the energy difference between $|\Phi_{i}\rangle$ and $|\Phi_{j}\rangle$, $\Delta_{ij}=E_{i}-E_{j}$, and $M_{ij}$ represents the transition matrix element of $e^{i\mathbf{q}\cdot \mathbf{r}}$ in three dimensions. In our model calculation, the electron is in the ground state along the transverse directions.  The transition element $M_{ij}$ thus takes the form of
\begin{align}
M_{ij}=\Pi\langle\Phi_{j}|e^{iq_{x}x}|\Phi_{i}\rangle,
\end{align}
with $\Pi$ being the average of $e^{i(q_{y}y+q_{z}z)}$ over the transverse directions, i.e., $\Pi=\langle e^{i(q_{y}y+q_{z}z)}\rangle$.
During this calculation, the wavefunction along the $z$ direction is truncated for the account of a finite length along the transverse direction.
Using three-dimensional phonon density of states, the relaxation rate can be written as
\begin{align}
\Gamma_{ i\rightarrow j}=\sum_{\lambda}\frac{1}{8\hbar\pi^{2}\rho c^{2}_{\lambda}}\oint q |(V^{\rm df}_{\mathbf{q},\lambda}-iV^{\rm pe}_{\mathbf{q},\lambda})M_{ij}|^{2} d \mathbf{q},
\end{align}
with the integral region satisfying the energy conservation condition $\hbar c_{\lambda}q=\Delta_{ij}$.  This result should be the most accurate when the nanowire is buried inside a substrate.  For a suspended nanowire, the relaxation rate should be further suppressed because of the reduced density of state for phonons.

In a cylindrical coordinate system, the relaxation rate caused by deformation potential $V^{\rm df}_{\mathbf{q}, \lambda} = D_{e} \delta_{\lambda,l}$ can be written as
\begin{align}
\Gamma^{\rm df}_{ i\rightarrow j}=\frac{\Delta_{ij}^{2}D^{2}_{e}}{8\hbar^{4}\pi^{2}\rho c^{5}_{l}}\int^{2\pi}_{0}\int^{\Delta_{ ij}}_{-\Delta_{ ij}}|M_{ij}|^{2}d\vartheta d\Delta^{zl}_{ ij},
\label{Gammade}
\end{align}
with $\Delta^{zl}_{ij}=\hbar c_{l}q_{z}$ and $\vartheta$ the azimuth angle.  Similarly, the relaxation rate caused by the piezoelectric interaction is
\begin{align}
\Gamma^{\rm pe}_{ i\rightarrow j}=\sum_{\lambda}\frac{\Delta^{2}_{ ij}e^{2}h^{2}_{14}}{8\hbar^{2}\pi^{2}\rho c^{3}_{\lambda}}\int^{2\pi}_{0}\int^{\Delta_{ ij}}_{-\Delta_{ ij}}|\Upsilon_{\lambda}M_{ij}|^{2}d\vartheta d\Delta^{z}_{ ij},
\label{Gammapi}
\end{align}
with
\begin{align}
&\Upsilon_{ l}=\delta_{\lambda,l}\frac{3(\Delta^{2}_{ ij}-\Delta^{z2}_{ ij})\Delta^{z}_{ ij}\cos\vartheta\sin\vartheta}{\Delta^{4}_{ ij}}, \nonumber \\
&\Upsilon_{ t1}=\delta_{\lambda,t}\frac{\Delta^{z2}_{ ij}\sqrt{\Delta^{2}_{ij}-\Delta^{z2}_{ ij}}\cos\vartheta-(\Delta^{2}_{ ij}-\Delta^{z2}_{ ij})^{3/2}\cos\vartheta\sin^{2}\vartheta}{\Delta^{3}_{ ij}\sqrt{\Delta^{z2}_{ ij}+(\Delta^{2}_{ ij}-\Delta^{z2}_{ ij})\sin^{2}\vartheta}}, \nonumber \\
&\Upsilon_{t2}=\delta_{\lambda,t}\frac{\Delta^{z}_{ ij}\sin\vartheta}{\cos\vartheta\Delta^{3}_{ ij}\sqrt{\varpi(1+\frac{\varpi}{\cos^{2}\vartheta(\Delta^{2}_{ij}-\Delta^{z2}_{ ij})})}}\{(\Delta^{2}_{ ij}-\Delta^{z2}_{ ij})(2\cos^{2}\vartheta-\sin^{2}\vartheta)-\Delta^{z2}_{ij}\}
\end{align}
where $\Delta^{z}_{ij}=\hbar c_{\lambda}q_{z}$ and $\varpi=\Delta^{z2}_{ij}+(\Delta^{2}_{ ij}-\Delta^{z2}_{ij})\sin^{2}\vartheta$.  The overall phonon-induced relaxation rate between states $|\Phi_{i}\rangle$ and $|\Phi_{j}\rangle$ is then
\begin{align}
\Gamma_{ i\rightarrow j}=\Gamma^{\rm df}_{ i\rightarrow j}+\Gamma^{\rm pe}_{ i\rightarrow j}.
\end{align}

\section*{Acknowledgements}
This work is supported by the National Key Research and Development Program of China (grant No. 2016YFA0301200), the NSFC (grant No. 11774022) and the NSAF (grant No. U1530401). R.L. is supported by the NSFC (grant No. 11404020) and Postdoctoral Science Foundation of China (grant No. 2014M560039).  X.H. acknowledges financial support by US ARO through grant W911NF1210609 and W911NF1710257 and thanks the CSRC for hospitality during the visit.

\section*{Author contributions statement}

Z.H.L. performed the derivations and numerical calculations under the guidance of J.Q.Y. and  X.H.. Also, R.L. participated in the discussions. All authors contributed to the interpretation of the work and the writing of the manuscript.

\section*{Additional information}

\textbf{Competing financial interests}: The authors declare no competing financial interests.

\end{document}